\documentclass[a4paper,11pt]{article}
\usepackage{pos}
\usepackage{amsmath}

\title{Dark-matter prolate halo shapes\\ from fits to SPARC galaxy rotation curves}
\ShortTitle{Prolate dark matter haloes}
\author{Adriana Bariego Quintana, Felipe J. Llanes-Estrada and Oliver Manzanilla Carretero}
\affiliation{Univ. Complutense de Madrid, Dept. F\'{\i}sica Te\'orica, Plaza de las Ciencias 1, 28040 Madrid, Spain}
\emailAdd{fllanes@fis.ucm.es}

\abstract{
The shape distortion of the presumed Milky Way dark matter halo can impact the local density of dark matter and thus the direct detection program.
We examine the population of galactic rotation curves measured by SPARC and fit them
to dark matter haloes that are distorted with a multipole density distribution, finding a significantly better fit with prolate haloes over spherically symmetric ones.
This is to be expected since the long-distance Rubin flattening $v(r)= {\rm constant}$ is the natural Kepler law due to a filamentary rather than a spherical source. Then, elongating the distribution brings about a smaller $\chi^2$, all other things being equal, including the 
use of several different radial dark matter profiles.  
The ellipticities that we fit to rotation curve data seem to be much more significant than those
computed in cosmological simulations of dark matter haloes. 
If the Milky Way's halo would be typical of the spiral-galaxy SPARC sample (which we presently ignore), the local dark matter density might currently be overestimated by a factor 2. This would carry on into dark-matter cross-section bounds.
}

\FullConference{%
  *** The European Physical Society Conference on High Energy Physics (EPS-HEP2021), ***\\
  *** 26-30 July 2021 ***\\
  *** Online conference, jointly organized by Universität Hamburg and the research center DESY ***
}

\begin{document}
\maketitle

\begin{figure}
\includegraphics[width=0.48\columnwidth]{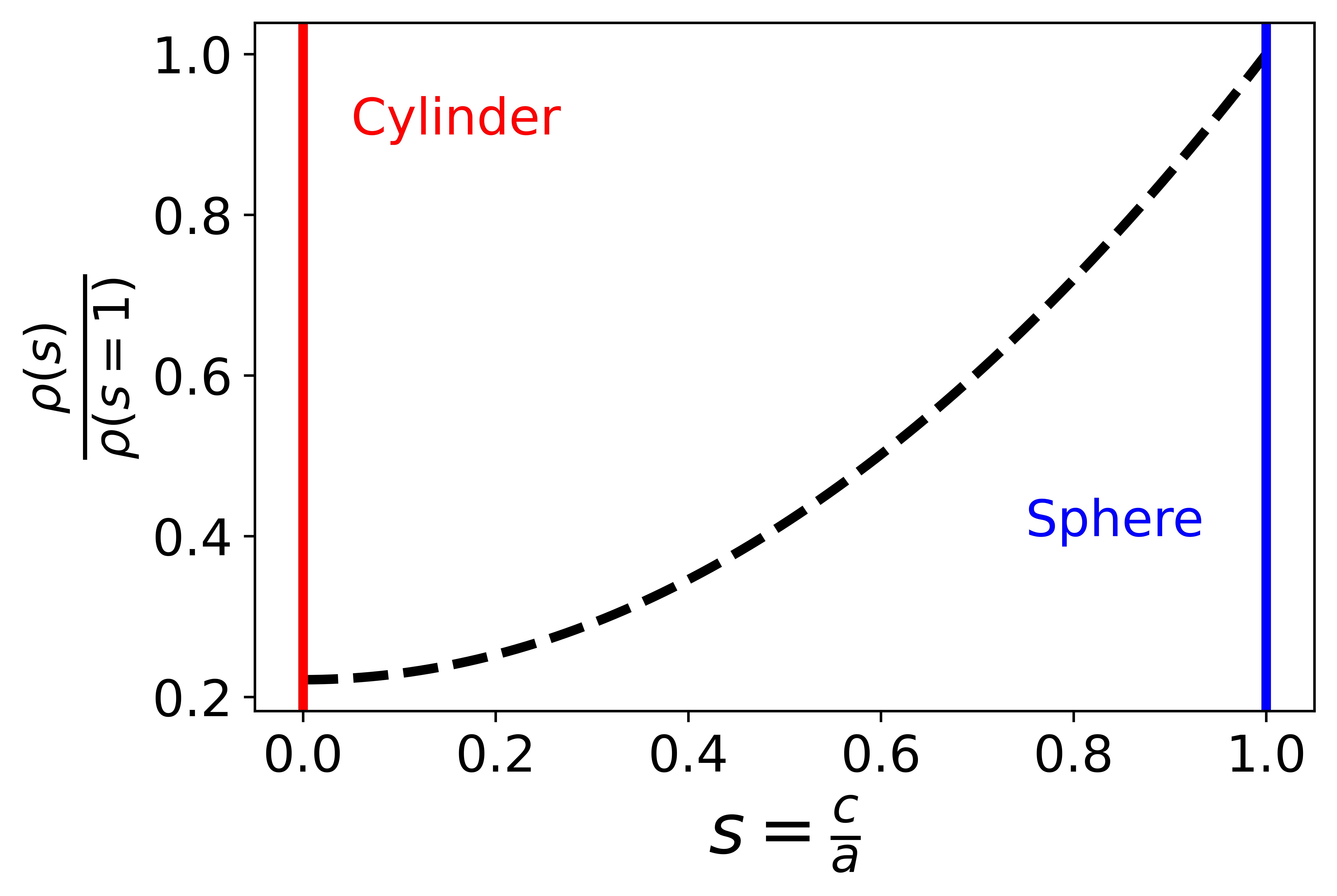}
\includegraphics[width=0.48\columnwidth]{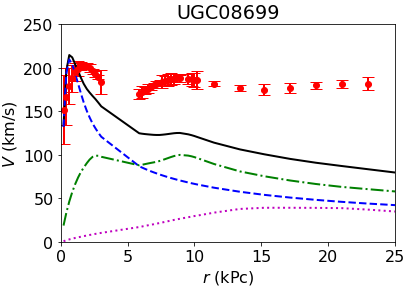}
\caption{{\bf Left:} Ratio of local dark matter density $\rho$ (at the position of Earth respect to the galactic center) of a prolate  ellipsoidal dark matter halo to the spherical one $\rho(1)$, as function of the deformation $s=c/a$, for a generalized logarithmic potential 
$\Phi \propto \log{\left[ 1 + R^{-2\alpha} \left (r^2 +s^2 z^2 \right)^\alpha \right] } $
{\bf Right:} Typical spiral galaxy rotation curve $v(r)$ from SPARC (data points with uncertainty bars) together with that predicted from the visible matter (also broken down by contributions as listed in Eq.~(\ref{contributions})). The difference between the top solid line and the data is attributed to either Dark Matter or Modified Gravity (MOND). From the database of such rotation curves we perform best fits to the shape of the galactic gravitational potential.
\label{fig:densitydrop}}
\end{figure}

Efforts towards dark matter (DM) direct detection~\cite{Billard:2021uyg}
require estimates of the density of presumed DM particles at Earth, currently to constrain DM scattering cross-sections off visible particles and for experimental planning.
A standard estimate  is $\rho_{DM\ Earth}\simeq 0.3$ GeV/cm$^3$ with typical uncertainty of order 50\%~\cite{Weber:2009pt}.
The shape of the DM halo can impact this density by a factor of order 0.1-1 (see figure~\ref{fig:densitydrop}, left plot). While it is not yet clear how the Milky Way's halo is shaped, we can turn to similar spiral galaxies for broader understanding.
Already computer simulations of dark matter~\cite{Allgood:2005eu} suggest a certain prolateness of the average galaxy, usually quoted, for an ellipsoid of revolution with equal minor half-axes $b=c$, in terms of the parameter $s:=c/a\simeq 0.6\pm 0.1$ taking the ratio to the major semiaxis. However other simulations also obtain a significant number of oblate haloes~\cite{Bett:2006zy}.

We here turn to the SPARC catalogue~\cite{Lelli:2016zqa} containing well measured galaxy rotation curves $v(r)$ and estimates of the visible, ordinary mass, assuming mass-to-light ratios respect to our Sun's of 0.7 (for the bulge) and 0.5 (for the disk).
Because different matter contributions in the galaxy additively contribute to the squared rotation velocity, we can subtract those estimates to obtain pseudodata $v^2_{DM}$ points  
that we fit to the DM gravitational potential, 
\begin{equation} \label{contributions}
v^2_{DM}  = v^2 - v^2_{\rm disk} - v^2_{\rm bulge} - v^2_{\rm gas}
\ \  \to \ \ v^2_{\rm DM}=-r \partial \Phi_{DM}/\partial r
\ .
\end{equation}

Two different analysis are here reported, independently carried out with separate computer codes,
both deploying the standard {\tt iminuit} minimization package~\cite{James:1975dr} to reach the minimum $\chi^2$ per degree of freedom for each geometry and each galaxy.


In our first analysis we compare the two extreme geometries of a purely cylindrical source~\cite{Llanes:2020} (which is the natural one to produce flat rotation curves $v\propto {\rm constant}$) and a purely spherical one (which is customarily assumed by analogy with other astrophysical objects and because of some cosmological simulations that do not discern halo shapes; this sphericity has already been challenged by Farrar and Loizeau~\cite{Loizeau:2021bum}).

Nine models, listed on the left half of table~\ref{tab:modelcomparison} show that indeed Newtonian physics with visible matter provide the worst fit as is well known, a consequence of $v_{\rm Kepler}(r)\propto 1/\sqrt{r}$ instead of a constant; also, we find that elongated geometries are as competitive as spherical DM densities, and that at fixed geometry, having more parameters at hand is preferable over a simpler model, because of the many data points, even considering the normalized $\chi^2_{\rm dof}$.  

\begin{table}
\caption{\label{tab:modelcomparison} {\bf Left:} Ordering of the best overall fit of the galaxy database for various models of DM with extreme spherical and cylindrical symmetry, and of two versions of Modified Newtonian Dynamics (our first analysis), according to the U-test of Wilcoxon-Mann-Whitney.\\ {\bf Right:} Average, uncertainty and median of the ellipticity $s=c/a$ over the SPARC galaxy sample showing that indeed prolateness is favored by flattening rotation-curve observational data (our second analysis). }
\begin{tabular}{|cc||ccc|} \hline
\# & Model approach & Ellipticity & (Woods-Saxon & profile) \\ \hline 
1 & Generalized logarithmic (cylindrical) potential & & & \\
2 & Einasto (spherical) profile & & & \\
3 & Woods-Saxon cylinder & $l$ included & $\langle s \rangle \pm \sigma_s $ & ${\rm Median}(s)$ \\
4 & Spherical pseudoisothermal halo & 0, 2 &  $0.45\pm 0.4$ & 0.29           \\
5 & MOND (simple) $a= \frac{1}{2} (a_N^2 +\sqrt{a_N^4+4a_N^2a_0^2})^{1/2}$ & 0, 2, 4 & 
$0.5\pm 0.6$ & 0.28 \\
6 & MOND (standard) $a= \frac{1}{2} (a_N +\sqrt{a_N^2+4a_Na_0})$ & & & \\
7/8 & Finite-width cylinder/ spherical Navarro-Frenk-White & & & \\
9 & Newtonian dynamics with visible matter only & & & \\
\hline
\end{tabular}
\end{table}

\begin{figure}
\includegraphics[width=0.57\columnwidth]{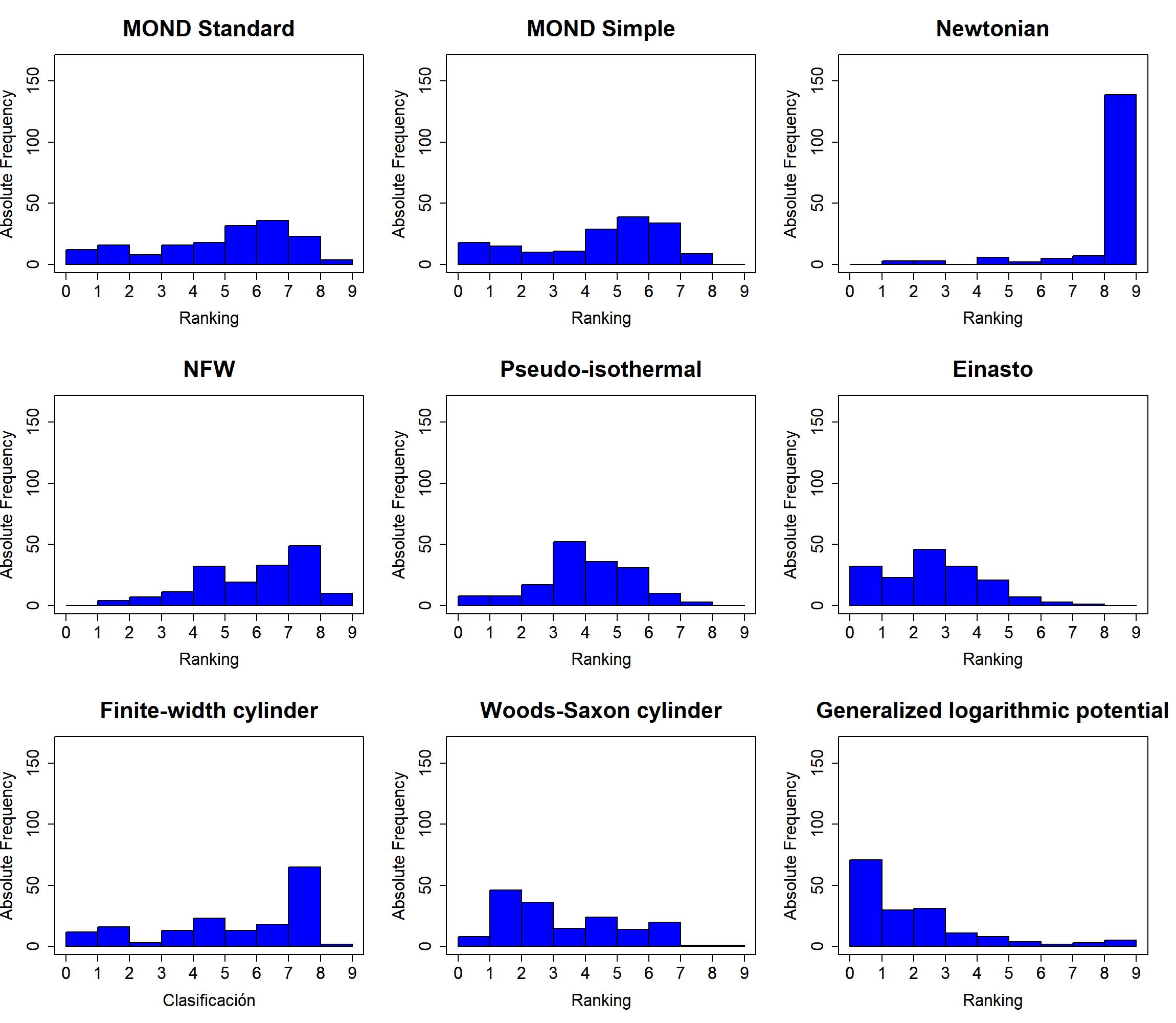}
\includegraphics[width=0.45\columnwidth]{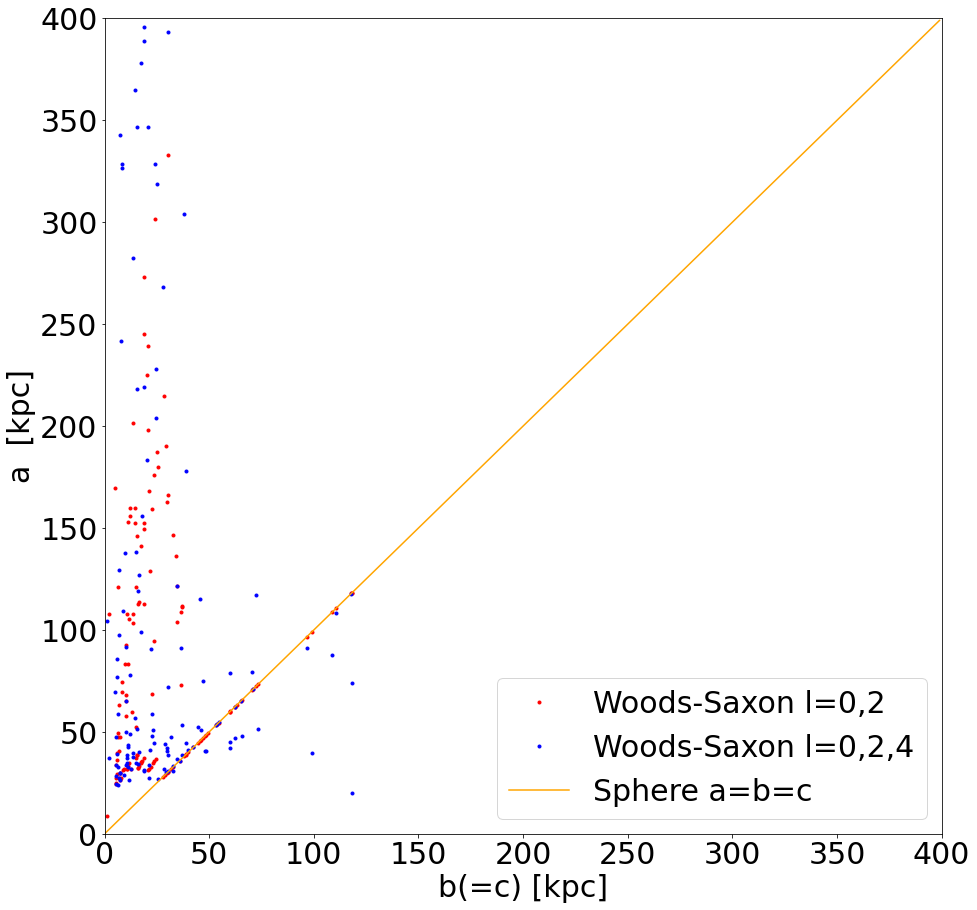}
\caption{\label{fig:modelsandpoles}
{\bf Left:} Histogram for each of the nine models of table~\ref{tab:modelcomparison} showing for how many individual galaxies each model provides the best fit, or the second best fit... down to the ninth and worst fit (basis of our first analysis).
{\bf Right:} Our second analysis shows that only a few galaxies in the SPARC database are best fit by an oblate gravitational potential (those below the $c=a$ bisectrix in the scatterplot), with the vast majority being prolate and some being extremely prolate, nearing cylindrical shape.}
\end{figure}

The left plot of figure~\ref{fig:modelsandpoles} is a histogram  counting the number of galaxies for which each one of the models is 1st, 2nd, 3rd, etc.; 
table 1 is based on that data.

For our second analysis a multipolar expansion of the gravitational potential is performed,
\begin{eqnarray}
    \frac{-\Phi(\textbf{r})}{4\pi G}&= & Y_{00}(\theta)\left[\frac{1}{r}\int^r_0 dr' \: (r')^2\rho_0(\textbf{r'})
\right]     
    + \frac{Y_{20}(\theta)}{5}\left[\frac{1}{r^3}\int^r_0 dr' \: (r')^4\rho_2(\textbf{r'})+r^2\int^R_r dr' \: (r')^{-1}\rho_2(\textbf{r'})\right]
\nonumber \\ 
    &+& \frac{Y_{40}(\theta)}{9}\left[\frac{1}{r^5}\int^r_0 dr' \: (r')^6\rho_4(\textbf{r'})+r^4\int^R_r dr' \: (r')^{-3}\rho_4(\textbf{r'})\right] + \dots
    \label{expansionmultipolar}
\end{eqnarray}
from which the rotation velocity  can be fit
once a model for the $\rho_i$ densities is adopted; we employ a step-function with its edge  
softened as the Woods-Saxon potential in nuclei, $\rho(r) = \frac{\rho_0}{1+e^{(r-R)/a_0}}$.

We find, in the right panel of table~\ref{tab:modelcomparison}, 
that the best fit potential is clearly prolate, even more so than in simulations. In fact, as the median and the actual scatter plot distribution (right panel of fig.~\ref{fig:modelsandpoles}) show, many spiral galaxies lean towards having an extremely prolate DM halo. 

Our conclusion is that ignoring the shape of the presumed DM halo entails an overestimation of the dark matter density at the arms of a typical spiral galaxy, such as at the Earth; statistically speaking, this can be a factor of 2 or larger (for a nonnegligible fraction of galaxies, up to 4).

\acknowledgments
This project has received funding from the spanish MICINN grants
PID2019-108655GB-I00, -106080GB-C21 and the U. Complutense de Madrid research group 910309 \& IPARCOS.

\newpage


\end{document}